\documentclass[conference]{IEEEtran}

\usepackage{color}
\usepackage{cases}
\usepackage{multirow}
\usepackage{graphicx,url}
\usepackage[linesnumbered,ruled,vlined]{algorithm2e}

\newcommand{\GF}[2]{\textcolor{black}{\textbf{}{#2}}}
\newcommand{\HT}[2]{\textcolor{black}{\textbf{}{#2}}}
\newcommand{\TM}[2]{\textcolor{black}{\textbf{}{#2}}}

\newtheorem{definition}{Definition}

\begin{document}
	
	\title{Sciunits: Reusable Research Objects}
	\author{\IEEEauthorblockN{Dai Hai Ton That, Gabriel Fils, Zhihao Yuan, Tanu Malik}
		\IEEEauthorblockA{School of Computing\\ 
			DePaul University, Chicago, IL, USA \\ 
			Email: \emph{\{dtonthat, gfils1, zhihao.yuan, tmalik1\}@depaul.edu} } }
	\maketitle
	\vspace{-4pt}
	\begin{abstract}
		Science is conducted collaboratively, often requiring 
		knowledge sharing about
		computational experiments. When experiments include only datasets, they can be shared using Uniform Resource Identifiers (URIs) or Digital Object Identifiers (DOIs). An experiment, however, seldom includes only datasets, but more often includes software, its past execution, provenance, and associated documentation. The Research Object has recently emerged as a comprehensive and systematic method for aggregation and identification of diverse elements of computational experiments. While a necessary method, mere aggregation is not sufficient for the sharing of computational experiments. Other users must be able to easily recompute on these shared research objects. In this paper, we present the \emph{sciunit}, a reusable research object in which aggregated content is recomputable. We describe a Git-like client that efficiently creates, stores, and repeats sciunits. We show through analysis that sciunits repeat computational experiments with minimal storage and processing overhead. Finally, we provide an overview of sharing and reproducible cyberinfrastructure based on sciunits gaining adoption in the domain of geosciences.
	\end{abstract}
	
	\IEEEpeerreviewmaketitle
	
	\section{Introduction}
	
	Research objects---aggregations of  digital artifacts such as code, data, scripts, and temporary experiment results---provide a means to share knowledge about computational experiments. In recent times, sharing computational experiments has become vital; scientific claims, inevitably asserted via computational experiments, remain poorly verified in text-based research papers. Research objects, together with the paper, provide an authoritative and far more complete record of a piece of research. 
	
	Several tools now exist to help authors create research objects from a variety of digital artifacts (see~\cite{Stodden2014} for several tools and~\cite{Malik:2014:SOLE} for a variety of research objects). The tools enable research objects to be shared on websites that disseminate scholarly information, such as Figshare~\cite{Figshare}. Despite their advantages, shared research objects do not permit easy reuse of their contents to verify their computations, or easy adaptation of their contents for reuse in new experiments. Often, the extent of reuse is subject to the amount of accompanying documentation, which may be limited to compilation and installation instructions. If documentation is scanty, research objects will remain unused.
	
	The minimum use case for sharing a computational experiment (in the form of a shared research object) involves repeating its original execution and verifying its results. To truly exploit its potential, however, it must support modified reuse. Therefore, the research object must be created and stored not as a simple aggregation of digital content, as previously advocated \cite{Belhajjame:RO:2015,RO-Bundle}, but in a readily-computable form: as a \emph{reusable} research object. We demonstrate the distinction in two ways.
	
	Consider a typical research paper with an analysis based on large amounts of code and data, and assume that the researcher authoring the paper has used the code and data to conduct a number of experiments that produce the paper's target figures and results. The example paper's digital artifacts relating to its experiments may be bundled together in a medium such as a file archive (.tar), compressed file format (.gz), virtual image, or container. A shared research object is free to use any of these mediums. A reusable research object, however, must use a virtual image or container, since it must produce a ``computational research object'' that, when downloaded and shared, will guarantee an instantly-executable unit of computation.
	
	Also consider the example paper's metadata, which, similar to the metadata in most papers, is interspersed throughout the project's written analysis, and throughout its code and data. The metadata can take many forms, including annotations, version information, and provenance. A shared research object's metadata usually serves a purely informational purpose, and is seldom used literally in the paper's experiments.  A reusable research object, however, utilizes literal metadata by directly linking it to the code and data of the experiments. In particular, provenance, if collected in standard form, can guide different forms of reusable analysis -- exact, partial, or modified reuse. Keywords and annotations can provide reference to additional datasets for modified reuse. In other words, a reusable research object can execute conditionally based on its embedded metadata, instead of simply including it as a stand-alone digital artifact that requires more interpretive labor to reason about and reuse. 
	
	In this paper, we describe the \emph{sciunit}, a reusable research object that has a lifetime beyond  being shared on scholarly exchange websites. The sciunit does not simply bundle digital artifacts, but uses application virtualization (AV) to automatically create a container of an executable application. In AV, operating system calls during application execution are interrupted to enable the copying of all binaries, data, and software dependencies into a container. The resulting container is portable and instantly reusable: it can be run on any compatible machine
	without installation, configuration, or root permissions.
	
	Similar to shared research objects, users can attach additional annotations to reusable research objects to describe containers. Each container also incorporates associated provenance, and users can use the included provenance to create 
	repurposed containers 
	These containers enable exact or partial repeatability of the sciunit.
	
	While AV facilitates the creation of reusable research objects, in its traditional form \cite{Guo:2011:CDEFull,Pham:2013:PTU} it is inherently inefficient when used to create multiple containers that are each based on slight modifications of an application. On repeated container creation, traditional AV methods behave the same regardless of the amount of similarity that exists between the original and modified applications. Traditional methods always create an entirely new container, which will contain wholesale duplication of digital artifacts, such as system dependencies, common binaries, or even common data files. Additionally, a large digital artifact present in two application versions, but that changes only slightly in content, will still consume its full amount of space in each corresponding container. Thus space consumption grows substantially, which is particularly of issue when a user shares different versions of an analysis or pipeline. We show how, when using our own AV method, multiple containers can be stored efficiently in one sciunit using a common block-based storage based on content de-duplication techniques~\cite{Muthitacharoen:ContentBaseddeDuplication}.
	
	We further increase reusability of sciunits by using their embedded metadata to help guide in their comprehension and modification. In particular, we use included provenance to provide an overview of the overall workflow of a container. When AV techniques are used to create a container, traditionally the collected provenance information is at the file and process level, which is too fine-grained to show the overall workflow. 
	
	This paper makes the following contributions: (i) 
	We present Sciunit-CLI\cite{SciunitCLI:2017:geotrusthub}, a Python/C-based Git-like client that creates sciunits, shares them on scholarly exchange websites such as Figshare and Hydroshare, and repeats shared sciunits either locally or remotely; (ii) We describe the AV method used in Sciunit-CLI to build a container for reuse; (iii) We describe versioned storage based on content deduplication methods to efficiently store multiple containers in a single sciunit; and (iv) We describe the interactive provenance visualization that summarizes embedded provenance in a container and simplifies repeating the container partially or modifying it.
	
	The rest of the paper is organized as follows:
	\begin{itemize}
		\item Section~\ref{sec:architecture}: overall architecture of our work.
		\item Section~\ref{sec:creation}: creating a \TM{}{sciunit} using application virtualization that builds a container with embedded provenance.
		\item Section~\ref{sec:storage}: storing multiple containers in a single sciunit.
		\item Section~\ref{sec:reuse}: utilizing provenance within a sciunit for repeating and reuse.
		\item Section~\ref{sec:graph visualization}: optimizing the embedded provenance for visualization in summarized graphs.
		\item Section~\ref{sec:experiments}: detailed experimental analysis.
		\item Section~\ref{sec:research object evolution}: evolution of research objects, and their creation and use in related applications.
		\item Section~\ref{sec:conclusion}: conclusions.
	\end{itemize}
	
	\section{The Sciunit-CLI: Architecture and use}
	\label{sec:architecture}
	
\begin{figure}[t!]
		\centering
		\includegraphics[width = 3in]{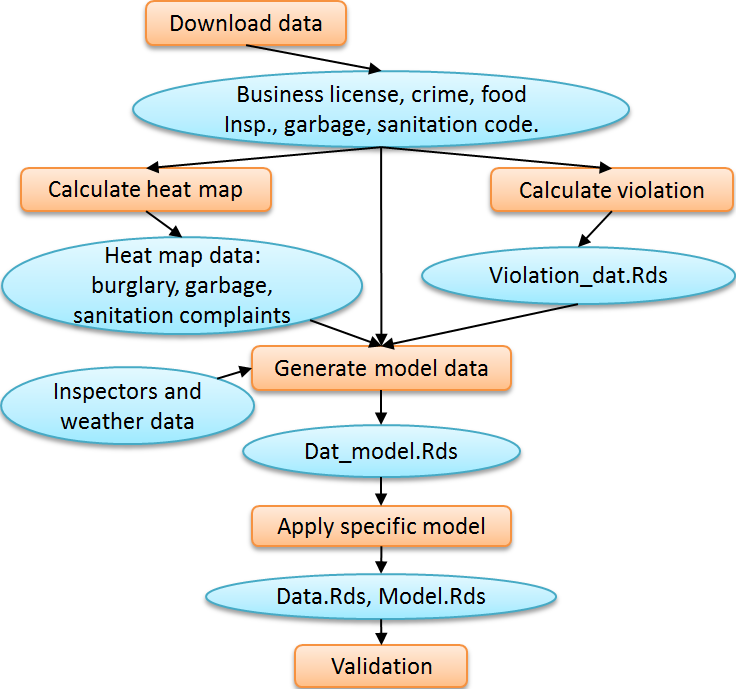}
		\caption{Conceptual view of the steps required to run the Food Inspection Evaluation \cite{FIE} predictive model}
		\label{fig:FIE workflow}
	\end{figure}

	Our reference implementation is a client program, the Sciunit-CLI\cite{SciunitCLI:2017:geotrusthub} that creates, stores, and executes reusable research objects.  We use a real-world example 
	to highlight the 
	primary commands and salient features
	of the client.
	Figure~\ref{fig:FIE workflow} shows an 
	example of 
	a predictive model used for forecasting critical violations during sanitation inspection \cite{FIE}. The software consists of scripts written in different languages (R and Shell) that operate on input datasets acquired from the City of Chicago Socrata data portal \cite{CoC}. The output of the predictive model is continually tested using a double-blind retrodiction; the Department of Public Health conducts inspections via its normal operational procedure, which are compared with the output of the model. The  pre-processing code is shared on GitHub~\cite{FIE-C},  the data is available via public repositories~\cite{CoC}, and the predictive model analysis is also published~\cite{FIE-P}.  
	
	Bundling these artifacts into a shared research object 
	would likely be inefficient given data from nine different sources, which changes periodically, making analysis conducted within a certain time range obsolete. The Sciunit-CLI can be used to build a reusable research object 
	consisting
	of identifiers of one or more re-executable containers, along with other listed digital artifacts. 
	
	
\begin{figure}[t!]
		\centering
		\includegraphics[width = 2.5in]{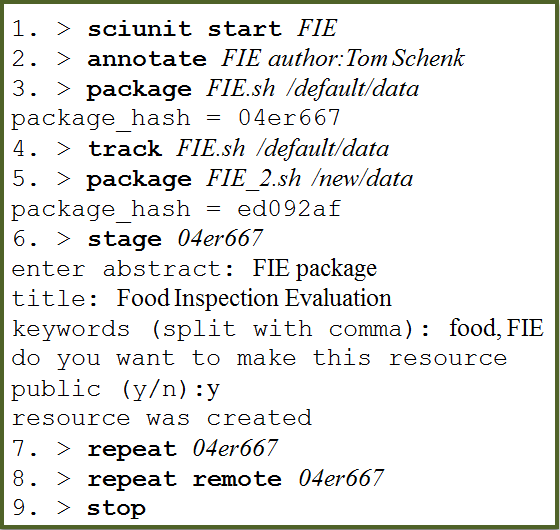}
		\caption{User interaction with sciunit client}
		\label{fig:client interaction}
	\end{figure}

	The Sciunit-CLI is a Git-like Python/C command-line interface (CLI) client used to build sciunits. Figure~\ref{fig:client interaction} shows a sample user interaction with this client. The user instantiates a namespaced sciunit titled \emph{\HT{myro}{FIE}} (Line\,1), and can associate annotations with the sciunit 
	(Line\,2). To create a container within the sciunit, 
	the user runs the application with the \emph{package} command (Line\,3). Packaging an application also incorporates provenance information of the application run; provenance can also be audited without creating a container (Line\,4). Many containers can be created within the same sciunit by using the \emph{package} command again (Line~5).
	
	The \emph{package} command makes minimal assumptions regarding the nature of the application. In particular, the user application can be written in any combination of programming languages, e.g. C, C++, Fortran, Shell, Java, R, Python, Julia, etc, or be used as part of a workflow system such as Galaxy~\cite{Goecks:2010:Galaxy}, Swift~\cite{Zhao:2007:Swift}, Kepler~\cite{Altintas:2006:Kepler} etc. While our description assumes local execution, in practice, an application's execution can be either local or distributed. We choose an example with local execution since the description of the underlying AV method (Section~\ref{sec:creation}) for distributed and parallel applications, such as database applications~\cite{Pham:ICDE:LDV} and HPC programs~\cite{Pham:2014:Thesis} is beyond the scope of the current paper. 
	
	
	
	The \HT{\emph{myro}}{}sciunit is stored locally unless explicitly shared with a remote repository
	using the \HT{\emph{publish}}{\emph{stage}} command, which instructs the client to upload the container to a Web-based repository (Line\,6). 
	The Sciunit-CLI uses Hydroshare~\cite{Hydroshare} for geoscience applications and Figshare~\cite{Figshare} otherwise as its Web-based repository.
	
	A container within the sciunit can be re-run directly 
	on the local machine with the \emph{repeat} command, or 
	on a remote execution server with the \emph{repeat remote} command 
	(Lines~7 and~8). 
	In our reference implementation, remote execution refers to execution via Hydroshare. On remote execution, the target container is automatically downloaded 
	to a remote execution server, and, if the container is compatible with the execution server's architecture, the execution server runs it and sends the results back to the user. 
	The user can also modify a container by downloading it, modifying its code or data and running it locally, and then uploading the modified container, at which point a new version of the container 
	can be staged.
	
	Further improvements of the \emph{repeat remote} command such as connecting with the remote server via \emph{ssh} or enabling partial remote executions is part of our ongoing work. The client, and accompanying server-side infrastructure that stores and manages sciunits, form a reproducible infrastructure, currently in use within the geosciences domain in the United States \url{(http://geotrusthub.org)}. The site provides 
	full technical documentation and examples from domain sciences using the client.

	\section{Creating sciunits}\label{sec:creation}
	
	Tools based on application virtualization method typically run in two modes: an audit mode to create a container, and an execution mode to re-run a container \cite{Guo:2011:CDEFull,Pham:2013:PTU}. In AV audit mode, a container of a user application is created as the user executes the application (in the context of auditing, such an execution is termed a \emph{reference execution}). We describe the audit process assuming that the application is running on a Linux machine. During execution, the Linux \emph{strace} utility is used to monitor the running application process. \emph{Strace} internally attaches itself to the process using the \emph{ptrace} system call to monitor all the system calls of the running process. It intercepts each system call\footnote{There are approximately 50 such calls defined in the POSIX standard} to determine the running process' state and the arguments to the system call.  For example, when a process accesses a file or a library using the system call \emph{fopen()}, the \emph{fopen()} call is intercepted. The intercepted system call is ``paused'' to examine input arguments and the process control block. For instance, in \emph{fopen()}, the file path parameter is extracted. By intercepting all calls, AV auditing determines all\footnote{Not all program dependencies can be detected through this method. But a program's static dependencies are much simpler to gather using programs such as \emph{file, ldd, strings, and objdump}. Our client provides commands for users to find additional dependencies, and include them, if necessary.} program binaries, libraries, scripts, and environment variables that a user program is dependent on. Inclusion of data files is optional, which the user may or may not want to package based on the size of the dataset. The audit process is similar for Windows and macOS, except that different OS-specific monitoring utilities are used.
	
\begin{figure}[t!]
		\centering
		\includegraphics[width = 2in]{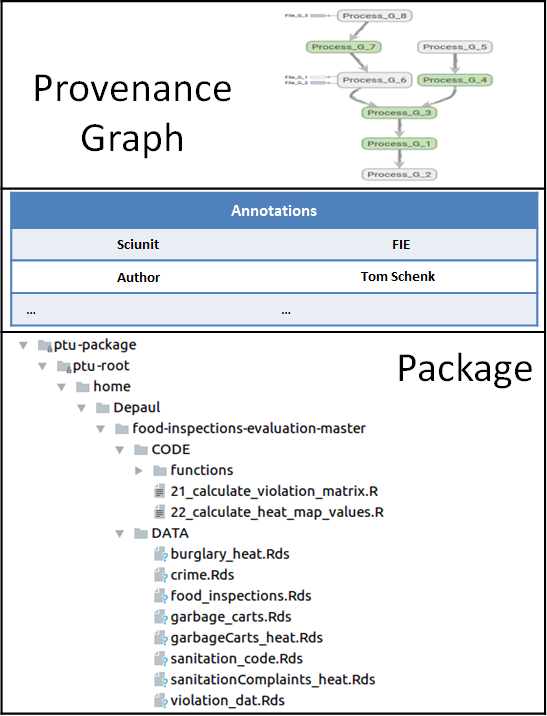}
		\caption{Example sciunit container}
		\label{fig:Package details}
	\end{figure}

	The system call pause time is brief, requiring only two lightweight context switches added to the normal system call flow; experiments show that the overhead of intercepting system calls is minimal. During the pause, the identified dependencies are used in two ways: first, to create a ``sandbox'' application container that includes all identified dependencies, and second, to create an interaction log of the reference execution. The sandbox container is named with a package hash and placed in a special ``root path'' (as described in Section \ref{sec:storage}), and contains all the dependencies that were identified during the reference execution audit. The dependencies are placed at the same path within the special root path as they were identified in the original system. Figure~\ref{fig:Package details} shows the contents of a container.\GF{ This path-mirroring has the side effect of exposing user directories and file system layout when the resulting container is reused. Thus, as a practice, creation of a reusable research object is best done within a shared or public namespace. }{}
	
	The interaction log generated during the AV audit phase contains interactions between processes when they are forked or execed, or between processes and files when files are opened or closed. 
	The log also 
	stores the logical range of times that processes interacted with other processes or with files. 
	A provenance graph is obtained by toplogically sorting the interaction log.
	
	
	
	In AV execution mode, the application is executed from the container itself by monitoring its processes with \emph{strace}, interrupting application system calls and extracting their path arguments, and redirecting all system call paths to paths within the special root path of the \HT{sandboxed}{sandbox} container. By redirecting all \HT{application}{}file requests into the container, the AV execution method fools the application program into believing that it is executing on the original audit-time machine with original file paths~\cite{Pham:2013:PTU}.
	
	The advantages of using the AV method are the ease with which a reusable research object can be created, and the machine-agnostic reuse that such an object provides. The disadvantages of the method are that the generated provenance is too fine-grained (at the file and process level) for ready analysis, and that repeated containerization can lead to many redundant files in the same research object. We address these two concerns in the next two sections.
	
	\section{Storing sciunits}\label{sec:storage}
	
	A reusable research object may include many containers. If the AV audit method described in Section~\ref{sec:creation} is used on an application to create a container for a sciunit, then each time that same application is audited all the same file dependencies of the application will be copied into a new container. This copying takes place even if the same dependencies were present in other previously-created containers based on the same audited application.
	One way to eliminate such dependencies is to check for duplicate dependencies as the container is created during the AV audit phase. 
	However, this slows the audit phase down delaying the construction of the container. The Sciunit-CLI de-duplicates on completion of the \emph{package} command as a background process.  
	
	Sciunit-CLI uses content-defined chunking 
	to divide the container's content 
	into small chunks identified by a hash value. 
	New chunks are compared to stored chunks, and whenever matches occur, redundant chunks are replaced with small references that point to stored chunks. 
	
	To identify chunks in a file, we do not use fixed-size chunking, which is simple and fast but faces the problem of low de-duplication
	ratio that stems from the boundary-shift problem \cite{Muthitacharoen:ContentBaseddeDuplication}. For example, if one or several bytes are inserted at the beginning of a file, all current chunk boundaries declared by FSC will be shifted and no duplicate chunks will be detected.
	Instead we use content-defined chunking (CDC)\cite{Muthitacharoen:ContentBaseddeDuplication}, that uses a sliding window technique on the content of files and computes a hash value (e.g., Rabin fingerprint~\cite{Rabin:fingerprinting}) of the window. In Rabin CDC, the Rabin hash for a window containing a $n$ byte sequence $B_1,B_2,\ldots,B_n$ is defined as a polynomial $RH(X_{(i,n)})\,=$: 
	\begin{equation}
	RH(B_1, B_2, ..., B_n)=\{\sum_{x=1}^{n}B_xp^{n-x}\}\,\bmod\,D
	\end{equation} in which D is the average chunk size. 
	Rabin hash is a rolling
	hash algorithm since it is able to compute the hash in an
	iterative fashion, i.e., the current hash can be incrementally
	computed from the previous value using a recurrence relation defined as:
	\begin{equation}
	RH(X_{(i,n)}) \leftarrow (RH(X_{(i-1,n)}) + X_i - X_{(i-n)})\,\bmod\,M 
	\end{equation}
	in which $n$ is the window size, $X_{(i,n)}$ represents the window bytes at byte position `$i$', and $M$ is the total length of the file. Using the recurrence relation, the hash value at \TM{}{any }byte position $i$ can be cheaply computed from the hash at byte position $i-1$.
	A chunk boundary is declared if the hash value satisfies some pre-defined condition, such as if the lowest $k$ bits of the Rabin hash value match a threshold value.

	Content-defined de-duplication is used in the popular Linux utility \emph{rsync} and we use it in a similar way in our work. However, unlike \emph{rsync} we search for hashes differently. In particular, instead of using a combination of fixed-size and rolling hashes, as used in \emph{rsync}, we simply iterate over all calculated hashes, speeding up  computation. 
	This is justifiable since we expect each research object to be fairly modest in size, unlike large-scale storage and backup systems where \emph{rsync} is commonly used.

\begin{figure}[t!]
		\centering
		\includegraphics[width = 3in]{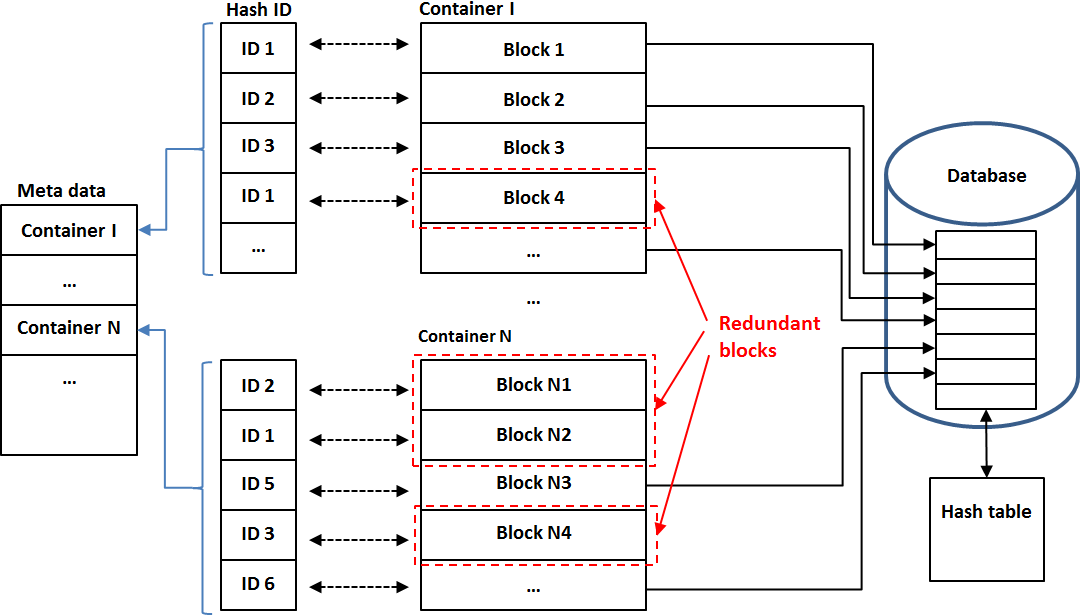}
		\caption{Block-based deduplication of containers}
		\label{fig:Block based deduplication}
	\end{figure}

	Once rolling hashes have been computed from a file, and a different block is detected, the difference itself can be be stored either as a delta or as a distinct block. The delta method is typically used when the predominant use case is to efficiently obtain a specific version of a file. In our case, we needed to strike a balance between storing multiple overlapping containers and storing versions of a single container. Thus we chose the distinct block method, as shown in Figure \ref{fig:Block based deduplication}, in which all unique blocks across all containers, versioned or not, are stored. \HT{Figure~\ref{fig:Block based deduplication} also shows a delta-based method may occupy more space than unique block based method, since it stores difference from the penultimate version.}{}
	
	Given this optimization, a container then is just a symbolic view over deduplicated storage, as shown in Figure \ref{fig:Block based deduplication}. However, for the user this use of optimized storage is opaque. 
	The Sciunit-CLI uses a manifest to store multiple containers, and to select a specific container to run.
	To run, the client first materializes the selected container by enumerating and simply concatenating the
	blocks corresponding to the selected container. The materialization requires negligible processing. 
	This procedure is fundamentally different from a delta-based mechanism, in which the blocks corresponding to a selected container will have to first reconstructed by applying the deltas. 
	
	\section{Reusing sciunits}\label{sec:reuse}
	
	When a sciunit is published, the server distinguishes between the computational part (i.e. the application container) and the non-computational part (i.e. the informational digital artifacts) of the sciunit. The computational part is associated with a cloud instance that can remotely execute the container on user request. A new user can reuse a published sciunit in one of three ways: (i) exact repeat-execution, (ii) partial repeat execution, or (iii) modified repeat execution. To exactly repeat, a container is simply downloaded and then run locally with the \emph{repeat} command, or run remotely with the \emph{repeat remote} command: the container will execute exactly as it did when it was created with the \emph{package} command. To partially repeat or run a modified repeat, a container is downloaded, processed for partial or modified execution, and then either run locally or published to run remotely. We now describe the processing required for partial and modified repeat executions in detail.
	
	\begin{algorithm}[t!]		        
		\DontPrintSemicolon
		\SetKwProg{Def}{}{:}{}                
		\Def{\textbf{BuildSubContainer}($selectedProcs$, $container$)} {
			$subContainer$ = initialize($container$)\;
			$allProcs$ = getAllProcs($container$)\;
			$requiredProcs$ = \textbf{getProcs}($selectedProcs$, $allProcs$)\;
			$reqProcDeps$ = \textbf{getDeps}($requiredProcs$)\;
			\ForEach{$dep$ \textnormal{\textbf{in}} \{$reqProcDeps$\}} {
				/* add dep to correct location in subContainer */\;
				add($dep$, $container$, $subContainer$)\;
			}
			\textbf{return} $subContainer$\;
		}
		\Def{\textbf{getProcs}($selectedProcs$, $allProcs$)} {
			$result$ = \{$selectedProcs$\}\;
			\ForEach{$proc$ \textnormal{\textbf{in}} \{$allProcs$\}} {
				\ForEach{$selProc$ \textnormal{\textbf{in}} \{$selectedProcs$\}} {
					\If{\textnormal{isDescendant}($proc$, $selProc$)} {
						$result$ = $result$ $\cup$ $proc$\;
						\textbf{break}\;
					}
				}				
			}
			\textbf{return} $result$\;
		}
		\Def{\textbf{getDeps}($requiredProcs$)} {
			$result$ = $\emptyset$\;
			\ForEach{$reqProc$ \textnormal{\textbf{in}} \{$requiredProcs$\}} {
				/* retrieve all related files and dependencies */\;
				$deps$ = relevantResources($reqProc$)\;
				$result$ = $result$ $\cup$ $deps$\;				
			}
			\textbf{return} $result$\;
		}
		\caption{Build sub-container for partial execution}
		\label{algorithm:build sub-container}
	\end{algorithm}
	
	\subsection{Partial Repeat Execution}\label{subsec:PartialRepeat}
	
	To partially repeat, a user selects one or multiple processes within a container. These processes are identified by their short pathname or PID, and the user can also use the provenance graph to aid in identification. While the provenance graph can be quite detailed for a user to choose specific processes, in Section \ref{sec:graph visualization} we describe how a user can see a summarized application workflow akin to the workflow presented in Figure \ref{fig:FIE workflow} from the provenance graph. Thus, for example, using the container from Figure 1, a user selects the processes ``Calculate violation" and ``Generate model data" as the group of processes to be partially repeated\HT{(starred in Figure \ref{fig:FIE workflow})}{}. Since this user-selected group of processes may not include all related processes needed for re-execution, we must determine these related processes, along with the data files they reference. The determined processes and files will constitute the new ``partial repeat" container or ``sub-container". Algorithm \ref{algorithm:build sub-container} shows the procedure for building the sub-container. It starts with the list of user-selected processes (\emph{selectedProcs}), and progresses to include all relevant processes and files by traversing the lineage of the graph (Lines 10-16). The \emph{getDeps} function assumes that any intermediate data files, if included as dependencies, still exist as generated from previous execution runs. The execution of this algorithm ensures that the data file ``Heat map data", generated from the previous run of the process ``Calculate heat map", is included in the sub-container, even though in the new partial repeat execution the process ``Calculate heat map" will not be re-executed.
	
	\subsection{Modified Repeat Execution}\label{subsec:ModifiedRepeat}
	
	To run a modified repeat of a sciunit container, a user examines a downloaded container and determines how particular computations within it should be modified (e.g. by modifying certain sections of code or input data). The sciunit's included provenance graph aids this modification task greatly. Next the user runs the modified container. To share the modification, the user would simply run it with the \emph{package} command, and then publish it with the \emph{stage} command. Enabling modification through a visualization mode, in which users can specify alternate processes or input data files assisted by a GUI, is part of future work.
	
	\section{Provenance graph visualization}\label{sec:graph visualization}
	
	Provenance information generated by AV audit methods is fine-grained. A graph created from a complete set of generated provenance, using normal visualization structures such as tree or list representations, would be far too replete to be of real practical value. When viewed, this graph would present significant system-level detail that would inhibit a basic comprehension of the overall application workflow. For example, the intuitive workflow of Figure \ref{fig:FIE workflow}, consisting of 12 nodes and 13 edges, would be represented fully as a dense graph of 146 nodes and 321 edges (Figure \ref{fig:Graph Optimization}(a) shows a part of this replete graph). Thus, to create a more intuitive graph, we use a graph summarization method that condenses the low-level details of the full generated provenance information. The graph summarization method is explained in detail in \cite{Li:2016:summarization}, and is briefly described in this section. We further describe how we extend the summarization method to create a graph that presents dynamic workflow cross-sections in a responsive visual interface.
	
	Given a directed graph $G=(V,E)$, where $V$ is the set of vertices\footnote{in our graph, a vertex is of type "file" or of type "process"} and $E$ is the set of edges, we denote $Input(u)$ and $Output(u)$ as the sets of input and output edges of vertex $u$. Respectively, $Input(u)=\{e|~\exists v \in V,~e=(v,u) \in E \}$, and $Output(u)=\{e|~\exists v \in V,~e=(u,v) \in E\}$. The direction of an edge characterizes the dependency of its vertices.  For example, a process $u$ spawned by process $v$ is represented by the edge $(u,v)$, and a file $u$ read by process $v$ is represented by the edge $(v,u)$. The graph $G$ is summarized based on the following two rules:
	
	\begin{figure*}[t]
		\centering
		\includegraphics[width = \linewidth]{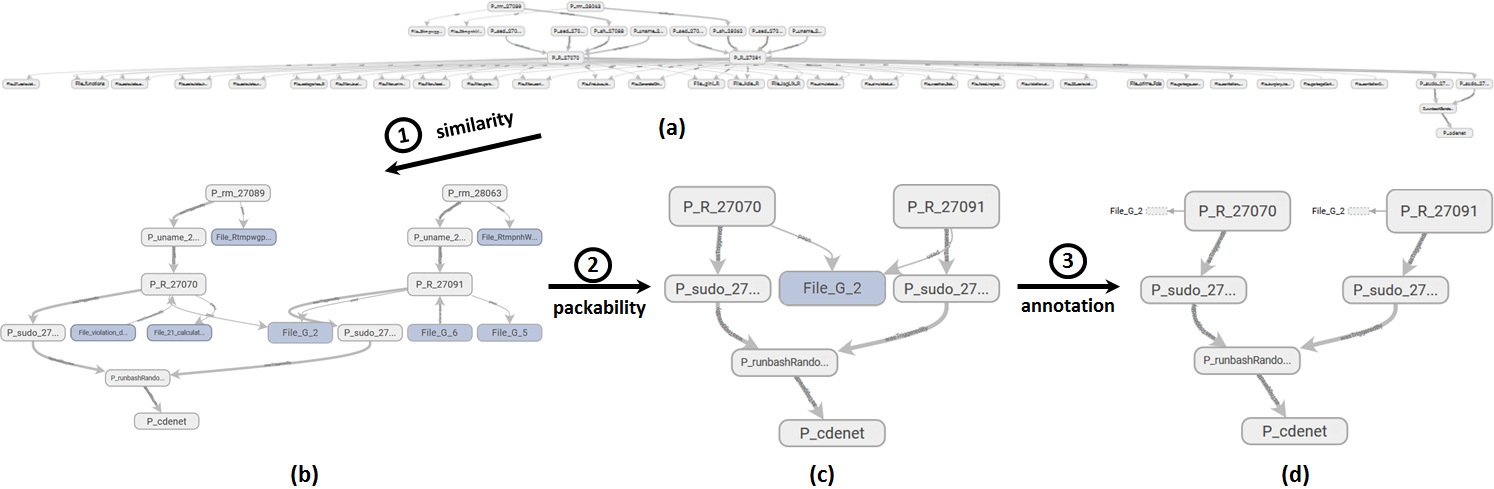}
		\caption{Graph summarization of a replete graph}
		\label{fig:Graph Optimization}
	\end{figure*}
	
	\begin{definition}[Similarity]
		Two vertices $u$ and $v$ are called \textit{similar} if and only if they share the same type and have the same input and output connection sets: $Type(u)=Type(v)$, $input(u)=input(v)$ and $output(u)=output(v)$.
	\end{definition}
	
	The similarity rule groups multiple vertices into a single vertex if the vertices have same type and are connected by the same number and type of edges. Additionally, edges of similar vertices will be grouped into a single corresponding edge. When applied to our provenance graph, this rule groups different files in the same directory.
	
	\begin{definition}[Packability]
		A vertex $u$ belongs to $v$'s $generalization~set$ if and only if vertex $u$ connects to $v$ and satisfies one of following conditions:
		\begin{itemize}
			\item Vertex $u$ is a file that has only one connection to process $v$: $Type(u)=file$ and \{$\exists ! e~|~e \in E \wedge (e=(u,v) \vee e =(v,u))$\}.
			\item Vertex $u$ is a process that has only one output connection to process $v$: $Type(u)=process$ and \{$\exists ! e~|~e \in E \wedge e = (u,v)$\}.
			\item Vertex $u$ is a file that has only two connections -- an output connection to process $v$ and an input connection another process $x$: $Type(u)=file$ and \{$\exists ! (e_1,e_2)~|~(\exists x \in V,~v \neq x)~ \wedge (e_1=(u,v) \in E, e_2=(\GF{k}{x},u) \in E)$\}.
		\end{itemize}
	\end{definition}
	
	The packability rule identifies hubs in the provenance graph by packing files or processes that are connected by single edges into their parent nodes. It also packs files that are generated by a single process and consumed by a single process into their parent processes by producing a process-to-process edge.
	
	\begin{figure}[h]
		\centering
		\includegraphics[width=2.75in]{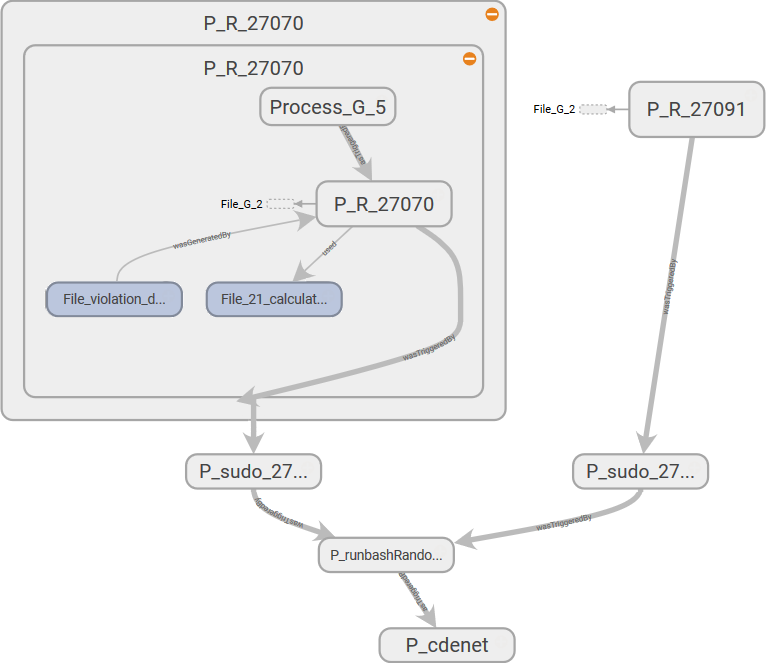}
		\caption{Expanded view of concealing node "P\_R\_27070"}
		\label{fig:Detailed View}
	\end{figure}
	
	When applied in sequence, the similarity and packability rules condense the detail-level of a graph while preserving its core workflow elements. Figure~\ref{fig:Graph Optimization} illustrates how applying these two rules to a replete graph produces a graph summary that shows the primary processes in a workflow. Figure \ref{fig:Graph Optimization}(a) presents the original replete provenance graph of one sub-task of the FIE workflow (the data processing steps ``Calculate Violation'' and ``Calculate Heat Map'' of Figure \ref{fig:FIE workflow}). Applying the two summarization rules produces the final graph in Figure \ref{fig:Graph Optimization}(c), which is similar to the conceptual workflow (except it is upside-down, due to the nature of provenance data flow).
	
	To lay out the summarized graph, we adopt two visualization techniques: scoping and annotation. In scoping, nodes similar to each other or packed together are represented as single nodes, which can be expanded on user action to reveal the details they conceal. For example, in Figure 7, similarity and packability rules group the nodes within the box into the single node ``P\_R\_27070" (process 27070 runs a subprocess using file ``21\_calulate\_violation\_matrix.R" and writes data to file "violation\_data.Rds"). The expanded view within the box was obtained by clicking on the concealing node ``P\_R\_27070." Here ``Process\_G\_5" is another concealing node hiding all the dependencies of the R process calculating the violation matrix.
	
	To further improve the layout of the graph, we use an annotation method that assigns higher visualization precedence to process nodes, but annotates them with corresponding file nodes. Figure~\ref{fig:Graph Optimization}(d) shows how the annotation ``File\_G\_2,'' which is a library dependency used both by ``P\_R\_27070'' and ``P\_R\_27091,'' is attached to the two process nodes that generated it. Thus, given a file with $n$ edges ($n \geq 2$), we replace this file with $n$ annotations. A user can always toggle the expanded view to see how the file and process nodes were originally connected. We choose to annotate files -- instead of processes -- since an application workflow is typically defined by the primary processes that it runs.
	
	\begin{figure}[h]
		\centering
		\includegraphics[width=\linewidth]{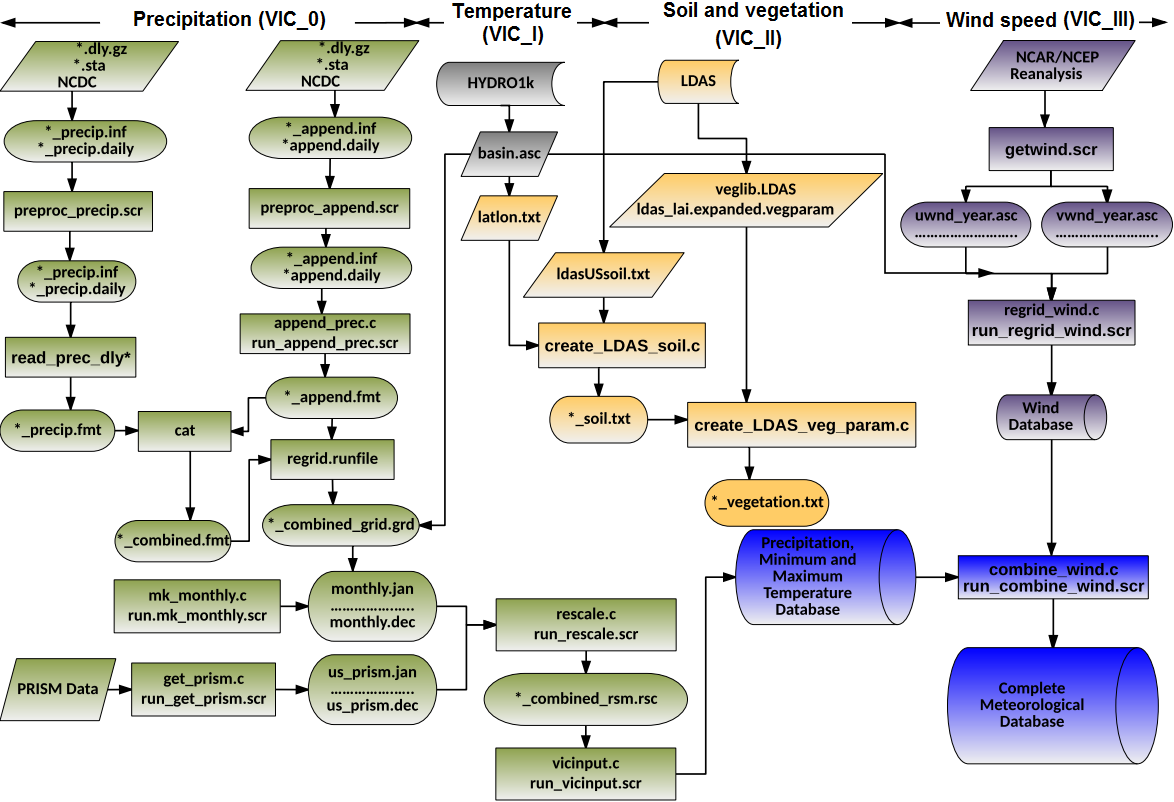}
		\caption{Conceptual view of the VIC workflow~\cite{billah2016using}}
		\label{fig:VIC workflow}
	\end{figure}
	
	\section{Experiments}\label{sec:experiments}
	
	The true usefulness of sciunits can only be measured by their adoption. Efficiency of creating sciunits can be a driving force in adopting the use of sciunits over traditional shared research objects. When an efficiently-versioned, easily-created sciunit is shared, along with an embedded, self-describing application workflow, we believe the probability for reuse will greatly increase. In this section, through two complex real-world workflows, we quantify the performance of packaging and repeating sciunits, the time and space overheads of storing them, and the efficiency of reusing them utilizing integrated provenance visualizations.
	
	\subsection{Use cases}\label{subsec:Experimental descriptions}
	
	We consider two real-world use cases for experimental evaluation: (i) the Food Inspection Evaluation ({\bf FIE}) \cite{FIE} workflow, a computationally-intense use case which has been the running example in our paper, and (ii) the Variable Infiltration Capacity ({\bf VIC}) \cite{billah2016using} model, an I/O-intensive data pre-processing pipeline for a hydrology model taken from \url{geotrusthub.org}. The first use case is notable for its transparency in its rigorous inspection audits, owing to the influence of the Open Data movement within the City of Chicago. The second use case is a highly-relevant test bed for sciunits: the VIC model is very popular in the hydrology community, and its data preprocessing pipeline, which relies heavily on legacy code, is notoriously difficult to reassemble \cite{billah2016using}.
	
	Tables \ref{table:fie application} and \ref{table:vic application} describe the details of FIE and VIC in terms of source code file programming languages, number of source code and data files, number of program files required as dependencies, and total application sizes (both FIE and VIC have four sub-tasks, labeled 0, I, II, and III, that are described below). Figures \ref{fig:FIE workflow} and \ref{fig:VIC workflow} show conceptual views of the application workflows for the two use cases. 
	
	Each use case is characterized by a shareable model, in which each step is conducted independently by one user, and subsequently shared with another user who builds upon or forks the shared workflow in the following step. Thus the FIE workflow, for example, is broken down into the following sub-tasks, each encapsulated in a single application: (i) FIE\_0, which calculates a heat map from downloaded inspection records; (ii) FIE\_I, which processes the heat map to generate data model inputs; (iii) FIE\_II, which applies a specific model and validates it; (iv) FIE\_III, which downloads the original inspection records and applies an end-to-end validation routine to the previous three sub-tasks. The download process of subtask iv is often the most time-consuming step. 
	
	The main sciunit client was implemented in Python and C. Sciunit's versioning tool was written in C++, using the block-based deduplication techniques proposed in \cite{Muthitacharoen:ContentBaseddeDuplication} and \cite{Rabin:fingerprinting}. Sciunit's provenance graph visualization was written in Python, using libraries from TensorBoard \cite{Tensorflow}. All sciunit client \emph{package} and \emph{repeat} experiments, along with their baseline normal application runs, were conducted on a laptop with an Intel Core i7-4750HQ 2.0 GHz CPU, 16 GB of main memory, and a 1 TB SATA SSD, running the Arch Linux 64-bit OS.
	
	\begin{table}[h!]
		\renewcommand{\arraystretch}{1.3}
		\caption{Food Inspection Evaluation Sub-task Applications}
		\label{table:fie application}
		\centering
		\begin{tabular}{|c|c|c|c|c|}
			\hline
			\textbf{} & \textbf{FIE\_0} & \textbf{FIE\_I} & \textbf{FIE\_II} & \textbf{FIE\_III}
			\\ \hline
			\textbf{Source Languages} & R, Bash & R, Bash & R, Bash & R, Bash
			\\ \hline
			\textbf{Source Files} & 19 & 20 & 24 & 29
			\\ \hline
			\textbf{Data Files} & 2 & 8 & 14 & 14
			\\ \hline
			\textbf{Dependency Files} & 255 & 255 & 411 & 659
			\\ \hline
			\textbf{Size of All Files} & 133.2MB & 178.4MB & 289.7MB & 306.6MB
			\\ \hline
			\textbf{Normal Run Time} & 52.046s & 238.833s & 295.785s & 7200s
			\\ \hline
		\end{tabular}
		\vspace{-10pt}
	\end{table}
	
	\begin{table}[h!]
		\renewcommand{\arraystretch}{1.3}
		\caption{Variable Infiltration Capacity Sub-task Applications}
		\label{table:vic application}
		\centering
		\begin{tabular}{|c|c|c|c|c|}
			\hline
			\textbf{} & \textbf{VIC\_0} & \textbf{VIC\_I} & \textbf{VIC\_II} & \textbf{VIC\_III}
			\\ \hline
			\textbf{Source Languages} & \multicolumn{4}{|c|}{C, C++, Python, C shell, Fortran}
			\\ \hline
			\textbf{Source files} & 35 & 61 & 77 & 97
			\\ \hline	
			\textbf{Data files} & 3689 & 6313 & 11460 & 11481
			\\ \hline
			\textbf{Dependency Files} & 247 & 260 & 314 & 357
			\\ \hline
			\textbf{Size of All Files} & 1.2GB & 1.3GB & 2.2GB & 2.3GB
			\\ \hline
			\textbf{Normal Run Time} & 158.734s & 306.069s & 363.147s & 377.29s
			\\ \hline
		\end{tabular}
		\vspace{-10pt}
	\end{table}
	
	\subsection{Creating Sciunits}
	\label{subsec:Reproducing overhead}
	
	Tables \ref{table:fie application} and \ref{table:vic application} present the baseline normal execution times for the sub-tasks of the two use cases. We note that each application encompasses substantial resources (in the form of code and data), has many external dependencies, and is also characterized by lengthy CPU-and-memory-intensive tasks. Additionally, the nature of FIE's processing tasks differ significantly from those of VIC. FIE front-loads its input data sets into memory, and then utilizes machine-learning logic to process its data. VIC also runs many intricate calculations, but differs from FIE in that it interlaces file input and output operations regularly throughout its code. This difference will be key in understanding that sciunits have minimal performance impact on most -- but not all -- types of applications.
	
\begin{figure}[h!]
		\centering
		\includegraphics[width=2.75in]{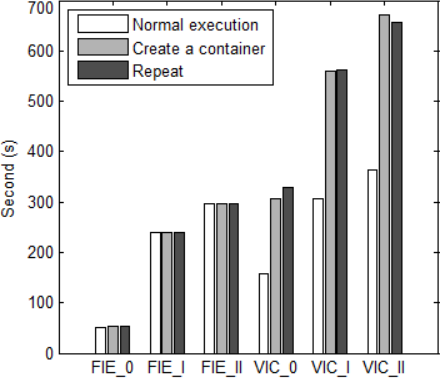}
		\caption{Execution times for normal runs, creating containers, and repeating}
		\label{fig:overhead}
	\end{figure}

	Figure \ref{fig:overhead} compares the baseline normal execution time of each subtask\HT{}{\footnote{Test results for the FIE\_III and VIC\_III sub-tasks were omitted due to significant amounts of network-dependent downloading operations.}} with the time consumed by packaging the sub-task with the sciunit \emph{package} command, and with the time consumed by repeating the sub-task with the sciunit \emph{repeat} command. We note that the performance impact of auditing and repeating on FIE's run times was negligible: auditing FIE with \emph{package} resulted in only a 3.6\% time increase, and executing FIE with \emph{repeat} added only a 1.3\% increase to run time. Conversely, both packaging and repeating VIC with sciunit each nearly doubled the original application run times: as noted in the preceding paragraph, it was evident that using sciunit with IO-intensive applications affected application performance significantly.
	
	We obtain one further observation from these experiments by comparing each application \emph{package} time with its corresponding \emph{repeat} time. Compared to application repeat increases, auditing increases were slightly higher. This difference can be understood by examining sciunit's behavior during AV audit-time: auditing entails copying an application's code and data into a sciunit container, but running the sciunit container with \emph{repeat}, however, only redirects to these copied files, and therefore precludes the file copy time.
	
	\subsection{Storing sciunits}\label{subsec:Versioning performance}
	
\begin{figure}[t]
		\centering
		\includegraphics[width=2in]{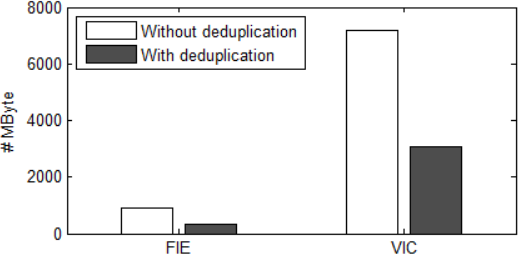}
		\caption{Saved space with content deduplication}
		\label{fig:Space Gain}
	\end{figure}

	Figure \ref{fig:Space Gain} presents the space saved by storing multiple sciunit containers in deduplicated storage. The total space consumed by FIE versions 0-III, storing each version separately, was 907MB, compared with a deduplicated total of 333MB. Similarly, VIC versions 0-III consumed 7GB in total separate storage, but when deduplicated consumed a total of 3GB.
	
	We also measured the computational complexity of committing and reconstructing a version to and from deduplicated storage. Committing a package involves taking an input container, constructing a single-file archive from it, and then performing deduplication on the archive against stored blocks. Consequently, commit times are a function of the size of the container. The reconstruction process only requires extracting the relevant blocks from storage and creating a package. Even though reconstruction is merely a block-concatenation process, it also entails recreating the original file entries from the block, and therefore can have a measurable time overhead.
	
\begin{figure}[t]
		\centering
		\includegraphics[width=3in]{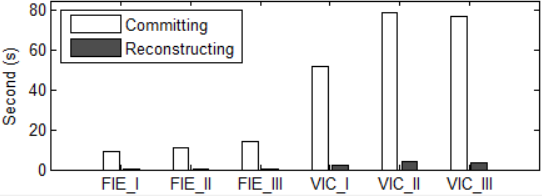}
		\caption{Execution times for committing and reconstructing versions}
		\label{fig:Execution Time}
	\end{figure}

	We measured both commit and reconstruction times that were far less than the normal baseline application execution times, and which would likely be imperceptible to users. Figure \ref{fig:Execution Time} shows the time in seconds for committing and reconstructing each sub-task\footnote{We depicted only packages of large size in order to clearly compare differences in commit and reconstruction times.}. Commit times were always greater than reconstruction times, due to the computation of rolling hashes during commit\footnote{The time for deduplication itself during commits was negligible, since it consists of single hash table lookups.}. Reconstruction times were dominated by the process of un-archiving individual files.
	
	\subsection{Reusing sciunits with Provenance Visualizations}\label{subsec:Graphic performance}
	
	Application virtualization has traditionally led to fine-grained provenance graphs that are often difficult to decipher. In this sub-section we determine if our summarization rules produce a usable provenance graph that is closer to a theoretical, intuitive user application workflow. We focus this discussion on experiments for the FIE sub-tasks, but note that experiment results for the VIC sub-tasks were similar.
	
\begin{figure}[t!]
		\centering
		\includegraphics[width=2.5in]{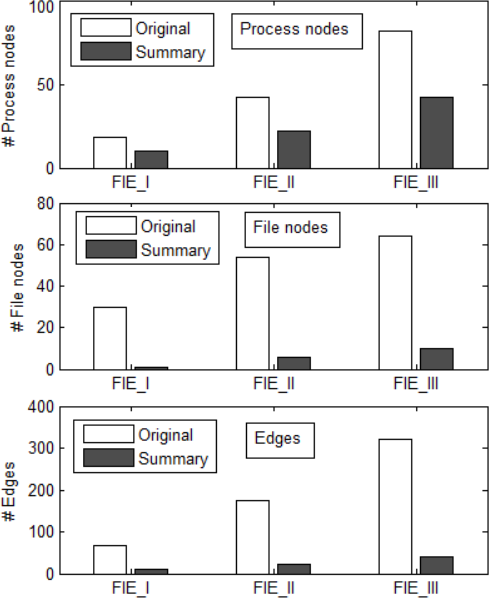}
		\caption{Number of nodes and edges in original and summarized graphs}
		\label{fig:Number Objects}
	\end{figure}

	To evaluate the effectiveness of summarization, we first considered three traditional, replete (i.e. fine-grained) provenance graphs generated by the legacy PTU application on auditing FIE\_I, FIE\_II, FIE\_III\footnote{We did not consider FIE\_0 in this analysis since its original replete graph was too small and simple to benefit measurably from summarization.}. We calculated the number of nodes (each a process or a file) and edges present in each replete graph. Next we calculated the number of nodes present in the corresponding sciunit container provenance graphs (these latter, summarized graphs were produced using the similarity and packability rules). Figure \ref{fig:Number Objects} depicts a comparison of the two graphs. It should be noted that the same provenance log (produced by PTU on audit) was used in the generation of both sets of graphs.
	
	Graph summarization reduced the number of file nodes, process nodes, and edges by averages of 90\%, 46\%, and 86\%. On closer examination, since the annotation technique only applies to files and their associated edges, we observed a larger decrease in the number of files and edges than the decrease in the number of processes. Of crucial -- but less measurable -- importance, we noted that the much smaller number of nodes and edges of the summarized graphs also carried more meaningful, intuitive labels, similar to those in Figure \ref{fig:Graph Optimization}.
	
	We also measured the number of clicks needed to expand summarized graphs to replete graphs. For FIE\_III, which had the largest graph, expanding any summarized node required a maximum of four user clicks to reach its replete view. Expanding all the nodes in this large graph took 45 clicks. This observation showed that graphs were  summarized very well spatially and intuitively, yet still capable of allowing fully-detailed provenance examination with a modest amount of user interaction.
	
	\section{\TM{Related work}{Evolution of Research Objects}}\label{sec:research object evolution}
	
	In this section we trace the evolution of the concept of research objects and their use toward advancing scholarly communication. 
	Research objects are increasingly seen as the new social object for advancing science \cite{DeRoure:DPRM:CRO}. They can be used for dissemination of scholarly work, measuring research impact, and assessing credit and attribution \cite{Yale:CiSE:2010}, which is the past was mostly done through research papers. The Research Object Model \cite{Belhajjame:RO:2015,Bechhofer:Linked:2013} is 
	a comprehensive standard defining the concept of a research object as a bundle of  artifacts that provides 
	a complete digital record of a piece of research. 
	Implementations of the standard \TM{has}{have} primarily focused on structured workflow objects \cite{Corcho:Workflow:2012} \cite{DeRoure:Preservation:2011} \cite{santana2015towards}, and have not yet encompassed general applications (i.e. applications executed without a formal workflow system). In this paper, we describe the sciunit client, a tool for creating a research object that includes containers created during run-time execution of an application, within  a workflow system \cite{Goecks:2010:Galaxy,Zhao:2007:Swift,Altintas:2006:Kepler} and outside \cite{Pham:2012:SOLE}.
	
	To create a research object, digital artifacts must be placed within it, either manually with explicit commands such as those used in RO-Manager \cite{RO-Manager} (a tool that uses the RO-Bundle specification \cite{RO-Bundle}), or automatically by using an AV tool such as Code, Data, and Environment (CDE) \cite{Guo:2011:CDEFull} \cite{Guo:2011:CDEShort} that containerizes an application as it executes. In this paper, we have chosen the AV tool Provenance-To-Use (PTU) \cite{Pham:2013:PTU} \cite{meng2015invariant}, which is built on top of CDE, to automatically capture provenance while creating containers, and have extended it for versioning and summarizing its provenance. We exclude more recent methods (such as 
	\cite{ivie2016prune}) that require users to learn new languages, and instead focus on the integration of DevOps tools in research objects.
	
	
	Recorded provenance can be made more conducive to new analyses by summarizing it using statistical \cite{MMS13} and non-statistical \cite{THP08} \cite{CCBD06} methods. Our sciunit client uses non-statistical methods 
	to summarize a research object's provenance, and extends the methods to visualize the summarized provenance spatially.
	
	Other methods to build and reuse containers, such as Topology and Orchestration Specification for Cloud Applications~\cite{standard2013topology}, still rely on user action to create the topology, relationship, and node specifications that are eventually translated to Dockerfiles \cite{qasha2016framework}. In our case, Docker is merely a wrapper for standardization, since application virtualization creates a self-contained container, and the translation to Dockerfiles from the collected provenance is fairly straightforward.
	
	\section{Conclusion}\label{sec:conclusion}
	
	Computational reproducibility \cite{Freire:2012} is a formidable goal requiring advancements in policy \cite{Victoria:2013:Toward}, user perception \cite{Penny:2016:Nature}, and reproducible practices and tools \cite{Stodden2014}. As we embrace this goal within the geosciences \cite{Malik:2017:geotrusthub}, we have encountered that general tools advocated for computational reproducibility must be enhanced in various ways. In this paper, we have challenged simple aggregation and advocated for containers, storing multiple of them with a relatively low storage cost, in logical sciunits, and their reuse in exact, partial, or modifiable forms using intuitive description of the reference execution. We demonstrated an easy-to-use Git-like client, the Sciunit-CLI that enables reproducibility for a wide variety of use cases. Yet, there are emerging requirements to address reproducibility within Jupyter notebooks, Matlab, distributed data-intensive programs, parallel HPC applications, which we hope to address as part of future work.

	
	\section*{Acknowledgment}
	
	The authors would like to thank Tom Schenk and Gene Leynes for the FIE use case and Jonathan Goodall and Bakinam Essawy for the VIC use case. The authors would also like to acknowledge support for this work from the National Science Foundation under grants NSF ICER-1639759, ICER-1661918, ICER-1440327, and ICER-1343816. Sciunit-CLI is downloadable from \cite{SciunitCLI:2017:geotrusthub}.
	
	\bibliographystyle{IEEEtran}
	\bibliography{./IEEEabrv,./sigproc}

\begin{thebibliography}{10}
\providecommand{\url}[1]{#1}
\csname url@samestyle\endcsname
\providecommand{\newblock}{\relax}
\providecommand{\bibinfo}[2]{#2}
\providecommand{\BIBentrySTDinterwordspacing}{\spaceskip=0pt\relax}
\providecommand{\BIBentryALTinterwordstretchfactor}{4}
\providecommand{\BIBentryALTinterwordspacing}{\spaceskip=\fontdimen2\font plus
\BIBentryALTinterwordstretchfactor\fontdimen3\font minus
  \fontdimen4\font\relax}
\providecommand{\BIBforeignlanguage}[2]{{%
\expandafter\ifx\csname l@#1\endcsname\relax
\typeout{** WARNING: IEEEtran.bst: No hyphenation pattern has been}%
\typeout{** loaded for the language `#1'. Using the pattern for}%
\typeout{** the default language instead.}%
\else
\language=\csname l@#1\endcsname
\fi
#2}}
\providecommand{\BIBdecl}{\relax}
\BIBdecl

\bibitem{Stodden2014}
V.~Stodden, F.~Leisch, and R.~D. Peng, Eds., \emph{{Implementing Reproducible
  Research}}.\hskip 1em plus 0.5em minus 0.4em\relax CRC Press, 2014.

\bibitem{Malik:2014:SOLE}
\BIBentryALTinterwordspacing
T.~Malik, Q.~Pham, and I.~T. Foster, ``{SOLE}: {T}owards {D}escriptive and
  {I}nteractive {P}ublications,'' \emph{Implementing Reproducible Research},
  vol.~33, 2014. [Online]. Available: \url{\url{https://osf.io/w6fp4/files/}}
\BIBentrySTDinterwordspacing

\bibitem{Figshare}
Figshare.com, ``Figshare,'' \url{https://figshare.com/}, 2017, [Online;
  accessed 2-May-2017].

\bibitem{Belhajjame:RO:2015}
K.~Belhajjame, J.~Zhao, D.~Garijo \emph{et~al.}, ``Using a suite of ontologies
  for preserving workflow-centric research objects,'' \emph{Web Semantics:
  Science, Services and Agents on the World Wide Web}, vol.~32, pp. 16--42,
  2015.

\bibitem{RO-Bundle}
S.~Soiland-Reyes, M.~Gamble, and R.~Haines, ``Research object bundle 1.0,''
  \url{https://researchobject.github.io/specifications/bundle/}, 2014, [Online;
  accessed 2-May-2017].

\bibitem{Guo:2011:CDEFull}
P.~J. Guo and D.~Engler, ``{CDE}: Using system call interposition to
  automatically create portable software packages,'' in \emph{USENIX}, 2011.

\bibitem{Pham:2013:PTU}
Q.~Pham, T.~Malik, and I.~Foster, ``Using {P}rovenance for {R}epeatability,''
  in \emph{TaPP}, 2013.

\bibitem{Muthitacharoen:ContentBaseddeDuplication}
A.~Muthitacharoen, B.~Chen, and D.~Mazi\`{e}res, ``A low-bandwidth network file
  system,'' \emph{ACM SIGOPS Oper. Syst. Rev.}, vol.~35, no.~5, pp. 174--187,
  2001.

\bibitem{SciunitCLI:2017:geotrusthub}
``{Sciunit-CLI},'' \url{https://bitbucket.org/geotrust/sciunit-cli.git}, 2017,
  [Online; accessed 10-Sep-2017].

\bibitem{FIE}
{City of Chicago}, ``{F}ood {I}nspection {E}valuation,''
  \url{https://chicago.github.io/food-inspections-evaluation/}, 2017, [Online;
  accessed 05-2017].

\bibitem{CoC}
------, ``Chicago data portal,'' \url{https://data.cityofchicago.org/}, 2017,
  [Online; accessed 7-May-2017].

\bibitem{FIE-C}
------, ``{F}ood {I}nspection {E}valuation predictions-source code,''
  \url{https://github.com/Chicago/food-inspections-evaluation}, 2016, [Online;
  accessed 7-May-2017].

\bibitem{FIE-P}
------, ``{F}ood {I}nspection {E}valuation,''
  \url{https://chicago.github.io/food-inspections-evaluation/predictions/},
  2017, [Accessed 05-2017].

\bibitem{Goecks:2010:Galaxy}
J.~Goecks, A.~Nekrutenko, and J.~Taylor, ``Galaxy: A comprehensive approach for
  supporting accessible, reproducible, and transparent computational research
  in the life sciences,'' \emph{Genome Biology}, vol.~11, 2010.

\bibitem{Zhao:2007:Swift}
Y.~Zhao, M.~Hategan \emph{et~al.}, ``Swift: Fast, reliable, loosely coupled
  parallel computation,'' in \emph{IEEE Congress on Services}, 2007.

\bibitem{Altintas:2006:Kepler}
I.~Altintas, O.~Barney, and E.~Jaeger-Frank, ``Provenance collection support in
  the kepler scientific workflow system,'' in \emph{IPAW}, 2006.

\bibitem{Pham:ICDE:LDV}
Q.~Pham, T.~Malik, B.~Glavic, and I.~Foster, ``{LDV}: Light-weight database
  virtualization,'' in \emph{ICDE}, 2015.

\bibitem{Pham:2014:Thesis}
Q.~Pham, ``A framework for reproducible computational research,'' Ph.D.
  dissertation, Dept. of Computer Science, University of Chicago, 2014.

\bibitem{Hydroshare}
Hydroshare.com, ``Hydroshare,'' \url{https://www.hydroshare.org/}, 2017,
  [Online; accessed 2-May-2017].

\bibitem{Rabin:fingerprinting}
M.~O. Rabin \emph{et~al.}, \emph{Fingerprinting by {R}andom
  {P}olynomials}.\hskip 1em plus 0.5em minus 0.4em\relax Center for Research in
  Computing Techn., Aiken Computation Lab, Univ., 1981.

\bibitem{Li:2016:summarization}
X.~Li, X.~Xu, and T.~Malik, ``Interactive provenance summaries for reproducible
  science,'' in \emph{IEEE e-Science'16}, Oct 2016, pp. 355--360.

\bibitem{billah2016using}
M.~M. Billah, J.~L. Goodall \emph{et~al.}, ``Using a data grid to automate data
  preparation pipelines required for regional-scale hydrologic modeling,''
  \emph{Environmental Modelling \& Software}, vol.~78, 2016.

\bibitem{Tensorflow}
Tensorflow.org, ``Tensorflow,'' \url{https://www.tensorflow.org/}, 2017,
  [Online; accessed 2-May-2017].

\bibitem{DeRoure:DPRM:CRO}
D.~De~Roure, ``{T}owards {C}omputational {R}esearch {O}bjects,'' in \emph{ACM
  Workshop on Digital Preservation of Research Methods and Artefacts}, 2013.

\bibitem{Yale:CiSE:2010}
{Yale Roundtable}, ``{R}eproducible {R}esearch,'' vol.~12, pp. 8--13, 2010.

\bibitem{Bechhofer:Linked:2013}
S.~Bechhofer, I.~Buchan, D.~De~Roure, P.~Missier \emph{et~al.}, ``Why linked
  data is not enough for scientists,'' \emph{Future Generation Computer
  Systems}, vol.~29, 2013.

\bibitem{Corcho:Workflow:2012}
O.~Corcho, D.~Garijo~Verdejo, K.~Belhajjame \emph{et~al.}, ``Workflow-centric
  research objects: First class citizens in scholarly discourse.'' 2012.

\bibitem{DeRoure:Preservation:2011}
D.~De~Roure, K.~Belhajjame, P.~Missier, J.~Gómez-Pérez \emph{et~al.},
  ``Towards the preservation of scientific workflows,'' in \emph{8th
  International Conference on Preservation of Digital Objects (iPRES)}, 2011.

\bibitem{santana2015towards}
I.~Santana-Perez and M.~S. P{\'e}rez-Hern{\'a}ndez, ``Towards reproducibility
  in scientific workflows: An infrastructure-based approach,'' \emph{Scientific
  Programming}, 2015.

\bibitem{Pham:2012:SOLE}
Q.~Pham, T.~Malik, I.~Foster \emph{et~al.}, ``{SOLE}: Linking research papers
  with science objects,'' in \emph{IPAW}, 2012.

\bibitem{RO-Manager}
\url{https://github.com/wf4ever/ro-manager}, 2016, [Accessed 02-May-2017].

\bibitem{Guo:2011:CDEShort}
P.~J. Guo, ``{CDE}: Run any {Linux} application on-demand without
  installation,'' in \emph{LISA}, 2011.

\bibitem{meng2015invariant}
H.~Meng, R.~Kommineni, Q.~Pham, R.~Gardner, T.~Malik, and D.~Thain, ``An
  invariant framework for conducting reproducible computational science,''
  \emph{Journal of Computational Science}, vol.~9, 2015.

\bibitem{ivie2016prune}
P.~Ivie and D.~Thain, ``Prune: A preserving run environment for reproducible
  scientific computing,'' in \emph{IEEE e-Science}, 2016.

\bibitem{MMS13}
P.~Macko, D.~Margo, and M.~Seltzer, ``Local clustering in provenance graphs,''
  in \emph{ACM CIKM}, 2013.

\bibitem{THP08}
Y.~Tian, R.~A. Hankins, and J.~M. Patel, ``Efficient aggregation for graph
  summarization,'' in \emph{ACM SIGMOD}, 2008.

\bibitem{CCBD06}
S.~Cohen, S.~Cohen-Boulakia, and S.~Davidson, ``Towards a model of provenance
  and user views in scientific workflows,'' in \emph{Data Integration in the
  Life Sciences}, 2006.

\bibitem{standard2013topology}
{Standard OASIS}, ``Topology and orchestration specification for cloud
  applications version 1.0,'' 2013.

\bibitem{qasha2016framework}
R.~Qasha, J.~Ca{\l}a, and P.~Watson, ``A framework for scientific workflow
  reproducibility in the cloud,'' in \emph{IEEE e-Science}, 2016.

\bibitem{Freire:2012}
J.~Freire, P.~Bonnet, and D.~Shasha, ``Computational reproducibility:
  State-of-the-art, challenges, and database research opportunities,'' in
  \emph{ACM SIGMOD}, 2012.

\bibitem{Victoria:2013:Toward}
V.~Stodden, P.~Guo, and Z.~Ma, ``Toward reproducible computational research: An
  empirical analysis of data and code policy adoption by journals,'' in
  \emph{PloS one}, vol.~8, 2013.

\bibitem{Penny:2016:Nature}
\BIBentryALTinterwordspacing
D.~Penny, ``{Nature Reproducibility survey},'' May 2016. [Online]. Available:
  \url{https://figshare.com/articles/Nature_Reproducibility_survey/3394951}
\BIBentrySTDinterwordspacing

\bibitem{Malik:2017:geotrusthub}
T.~Malik, ``{GeotrustHub},'' \url{https://geotrusthub.org/}, 2017, [Online;
  accessed 10-Sep-2017].

\end{thebibliography}
	
\end{document}